\documentclass{article}
\usepackage{comment} 
\usepackage{fullpage} 
\usepackage{graphicx}
\usepackage{etoolbox}   
\usepackage{verbatim}   
\usepackage[dvipsnames]{xcolor}
\usepackage{fancyvrb}
\usepackage{hyperref}
\usepackage{menukeys}
\usepackage{titlesec}
\usepackage{dcolumn}
\usepackage{bm}

\usepackage{xcolor}

\title{A Josephson Parametric Oscillator-Based Ising Machine}
\author{S. Razmkhah\thanks{Ming Hsieh Department of Electrical and Computer Engineering, University of Southern California, Los Angeles, CA, USA.}, M. Kamal\textsuperscript{*}, N. Yoshikawa\thanks{Department of Electrical and Computer Engineering, Yokohama National University, Yokohama, Japan.}, M. Pedram\textsuperscript{*}}%

\date{\today}
\titlespacing*{\section}{0pt}{5.5ex plus 1ex minus .2ex}{2ex}
\titlespacing*{\subsection}{0pt}{3ex}{2ex}

\begin{document}
\maketitle


\begin{abstract}
\noindent
Ising machines (IMs) have emerged as a promising solution for rapidly solving NP-complete combinatorial optimization problems, surpassing the capabilities of traditional computing methods. By efficiently determining the ground state of the Hamiltonian during the annealing process, IMs can effectively complement CPUs in tackling optimization challenges.
To realize these IMs, a bi-stable oscillator is essential to emulate the atomic spins and interactions of the Ising model. This study introduces a Josephson parametric oscillator (JPO)-based tile structure, serving as a fundamental unit for scalable superconductor-based IMs. Leveraging the bi-stable nature of JPOs, which are superconductor-based oscillators, the proposed machine can operate at frequencies of $\sim$7.5GHz while consuming significantly less power (by three orders of magnitude) than CMOS-based systems. Furthermore, the compatibility of the proposed tile structure with the Lechner-Hauke-Zoller (LHZ) architecture ensures its viability for large-scale integration.
We conducted simulations of the tile in a noisy environment to validate its functionality. We verified its operational characteristics by comparing the results with the analytical solution of its Hamiltonian model. This verification demonstrates the feasibility and effectiveness of the JPO-based tile in implementing IMs, opening new avenues for efficient and scalable combinatorial optimization in quantum computing.
\end{abstract}

\maketitle
\section{\label{sec:Intro}Introduction\protect\\}
\noindent
As semiconductor fabrication technology approaches its physical limits, the conventional scaling of semiconductors, which has driven the growth in integration density and operating speed of CMOS-based designs, is reaching its end \cite{ref1,ref2}. Therefore, exploring alternative technologies and architectures beyond CMOS that can enhance systems' performance and energy efficiency is essential. One solution to improve the efficiency of conventional computers is developing accelerator platforms for today's computing-intensive applications, such as artificial intelligence (AI), optimizations, pattern recognition, bio-informatics, and novel material simulations, that can keep up with the growing complexity of these problems.

Many of the computational problems encountered in current applications become increasingly intractable as the problem size scales up. This exponential increase in computational resources and time needed to solve such problems poses a significant challenge \cite{ref3}. Combinatorial optimization problems are often known to be nondeterministic polynomials (NP-complete), implying that finding the exact optimal solution can require exponential time in the worst-case scenario.

A promising approach for solving NP-complete problems involves identifying physical phenomena that can be modeled as a combinatorial problem. By mapping the original problem onto a physical system, we can leverage the system's inherent properties and dynamics to solve the problem. As the system reaches a stable state, the outcome represents the solution to the initial combinatorial optimization problem and solves the problem efficiently. 

An intriguing physical system with a combinatorial model is the behavior of free electrons' spins within a material's lattice. The material's lattice vibrates as it heats up, causing the electrons to move more freely. As the material cools down, the interaction force between electrons and the lattice exceeds the electrons' kinetic energy. Consequently, the electrons settle into a configuration that minimizes the lattice's energy. This phenomenon was initially mathematically described by Ernst Ising \cite{ref4}.
In the Ising model, each spin can either be up or down. The spin's behavior in the Ising model is determined by an energy function that relies on the states of the neighboring spins known as the Ising Hamiltonian. The electron configuration in the minimum value of this function determines the state of the matter.

Solving NP-complete problems using the Ising Model (IM) can offer a significant speedup compared to conventional computers \cite{ref5}. Thus, mapping NP-complete problems onto the Ising Model is advantageous, as finding the solution is equivalent to finding the ground state of the system's Hamiltonian. Additionally, IMs are highly effective in formulating and solving quadratic unconstrained binary optimization (QUBO) problems \cite{ref6}. This makes them a powerful tool for addressing a variety of optimization challenges.

IMs can be implemented using various interacting bistable oscillators as basic building blocks, such as optical parametric oscillators (OPO) \cite{ref7,ref8}, CMOS-based electronic oscillators \cite{ref9,ref10}, spintronics and magnetic systems \cite{ref11}, and quantum annealers \cite{ref12}.

While optical systems offer notable benefits for implementing Ising model solvers, including high processing speed and minimal noise levels, they face challenges related to integration limitations, bulkiness, and the need for long optical fibers \cite{ref10}. 
Another approach involves CMOS-based implementations for realizing the Ising model \cite{ref13,ref14,ref15}. Although CMOS-based implementations offer simplicity through techniques like iterative annealing in memory (AIM), they still encounter delay and energy consumption challenges.

Superconductor electronics (SCE) offer an alternative to semiconductors, providing higher computing speed and reduced power consumption \cite{ref16,ref17,ref18}. Quantum annealers, like DW2Q \cite{ref19}, utilize superconducting qubits representing electron spins, enabling high-speed operations through magnetic flux interactions. However, quantum annealers face challenges related to the sparse "Chimera" coupling graph architecture, which necessitates problem modifications and minor embedding methods \cite{ref20}. Additionally, achieving sub-Kelvin temperatures for observing quantum states adds complexity and costs to these systems.

This paper introduces a practical superconductor-based IM system that operates at 4.2K. A key component of our approach is the utilization of Josephson parametric oscillators (JPOs) \cite{ref21} as the fundamental spin elements within the IM. We propose a JPO structure that enables the formation of a four-body interaction network, a tile (plaquette), composed of six JPOs. 
The tile represents the fundamental building block in the IM architecture based on the LHZ (Lechner, Hauke, and Zoller) approach \cite{ref22}. 
Through inductive coupling, we establish controllable interactions between the JPOs by manipulating the phase differences between pairs of JPOs. These interactions drive convergence towards the minimum energy state of the Hamiltonian during the annealing process.
\section{\label{sec:Theory}Ising Machine\protect\\}
\noindent
An Ising machine determines the configuration states that minimize the interaction energy between spins. In this context, the Ising Hamiltonian for $N$ spins can be expressed as,

\begin{equation}
\mathcal{H} = -\sum_{i=1}^{N}h_i\sigma_i - \sum_{j,i}^{N}J_{ij}\sigma_i\sigma_j
\label{eq:ising_hamiltonian}
\end{equation}
where $\sigma_i$ ($\in {-1,+1}$) is the $i^{th}$ spin, $J_{ij}$ is the coupling interaction between the $i^{th}$ and $j^{th}$ spins, and $h_i$ is the local field of the $i^{th}$ spin.

The IM can solve the QUBO problem by mapping it onto the Ising Hamiltonian. This involves substituting the variable $x$ ($\in {0,1}$) in the QUBO with $\frac{1}{2}(\sigma-1)$ in the Ising Hamiltonian. By finding the ground state of $\mathcal{H}$, IM can effectively solve NP-complete combinatorial problems. However, achieving universal annealing requires control over each two-body interaction ($J_{ij}$). The challenge of physically implementing the IM for large-scale problems arises from the all-to-all interactions among spins represented by the interaction matrix $J$. The LHZ IM was introduced to address scalability in \cite{ref23}. The LHZ architecture offers a compact solution by eliminating the need for all-to-all interactions between spins.

In the LHZ IM, logical bits $\sigma_i$ define the Ising Hamiltonian (\ref{eq:ising_hamiltonian}), while physical bits $\tilde{\sigma_i}$ represent the relative configuration of two logical bits along a specific connection edge $J_{ij}$. Parallel and anti-parallel alignments correspond to 1 and 0, respectively. By treating optimization parameters $J_{ij}$ as local magnetic fields, the LHZ structure achieves full programmability with local control. The system size expands from $N$ logical bits to $K = N(N-1)/2$ physical bits to accommodate all interaction matrix elements.

\begin{figure}[h]
\centering
\includegraphics[width=0.4\linewidth]{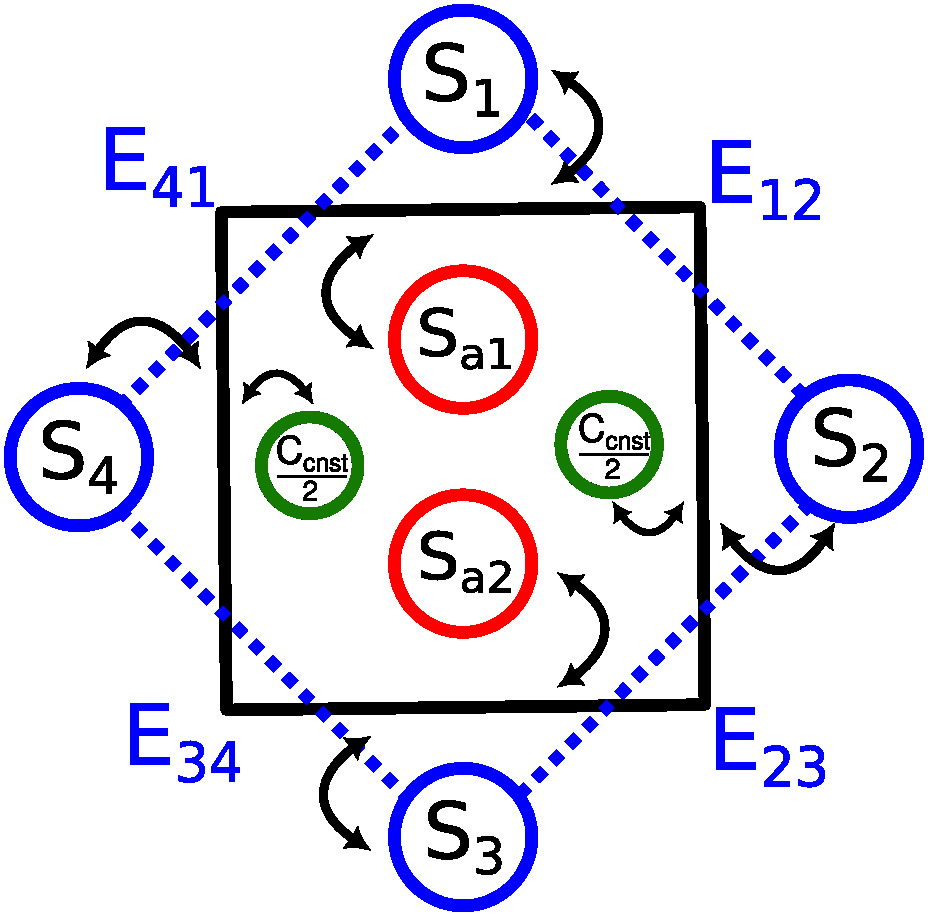}
\caption{A general structure of the JPO-based tile. $S_1$ to $S_4$ are the JPO-based spin nodes that are coupled together, and $S_{a1}$ and $S_{a2}$ are JPO-based ancilla spins which are used to impose the constraint $C_l$. $C_{cnst}$ is the offset value added to ensure that $C_l$ is always negative in the tile structure.}
\label{fig10}
\end{figure}

The physical bits form a 2-D triangular configuration, with $N-1$ bits in the base and decreasing levels above, ending in a single bit at the apex. An additional row of $N-2$ fixed physical qubits with a spin value of '1' completes the architecture. Each tile represents a four-spin interaction pattern. Fig.~\ref{fig10} illustrates an example tile structure employed within the LHZ architecture with four logical JPOs, and two ancillary JPOs enforcing the constraints $C_l$. The constant term is added to ensure term $C_l$ is always negative.

The increased degrees of freedom is balanced by $K-N+1$ constraints $C_l$, ensuring an even number of spin-flips along any closed loop in the logical bit. These constraints can be implemented using local interactions in simple square-lattice geometry. Consequently, the optimization problem is encoded as,
\begin{equation}
\mathcal{H} = \sum_{k=1}^{K} J_{k}\tilde{\sigma_i} + \sum_{l=1}^{K-N+1} C_l
\label{eq:LHZ_hamiltonian}
\end{equation}
The vector $J_k$ encompasses all $K$ elements of the interaction matrix $J_{ij}$, effectively converting the optimization parameters into local fields that can be easily controlled and applied to the physical bits. Furthermore, the constraint $C_l$ associated with the $l^{th}$ tile can be defined as,
\begin{equation}
C_{l} = -C \tilde{\sigma}^{(l,n)} \tilde{\sigma}^{(l,s)} \tilde{\sigma}^{(l,w)} \tilde{\sigma}^{(l,e)}
\label{eq:constraints}
\end{equation}

This equation represents the 4-body interaction between the physical bits (the north, east, south, and west) within the tile, with a penalty parameter of $C$. By applying the annealing process and reaching the ground state of $\mathcal{H}$, the solution to the optimization problem can be determined (except for a global inversion) by appropriately reading out a selection of $N-1$ physical bits from the total of $N(N-1)/2$ available.
\section{\label{sec:Method}Superconductor-based Components\protect\\}
\subsection{Josephson Parametric Oscillator (JPO)}
\noindent
Superconductor circuits offer unique quantum behavior at a macro scale, making them well-suited for innovative circuitry. Conduction in these circuits is facilitated by super-currents from Cooper pairs that act as quasiboson particles. Like photons in optical circuits, Cooper pairs possess the same wave function. However, unlike photons, Cooper pairs extensively interact due to their electric charge. This distinction enables the creation of bistable parametric oscillators, similar to OPOs, using superconductor circuits. These oscillators can be coupled and scaled up, integrating numerous oscillators on a single chip.

At the heart of most superconductor circuits, including JPOs, lies the Josephson junction (JJ). A JJ forms at weak links within a superconductor, generating a constant current due to the phase difference in the wave function at the junction's terminals. The behavior of super-currents in JJs is described by the Josephson DC and AC equations \cite{ref24}.

Considering the Josephson equation pair, it is evident that the JJ possesses an inductance that relies on the magnetic flux or the current passing through it. This inductance can be determined by equation \ref{eq:1} as,
\begin{equation}
L_{JJ} = \frac{\hbar}{2eI_{C}} = \frac{\Phi_0}{2\pi I_{C}},
\label{eq:1}
\end{equation}
where $L_{JJ}$ is the inductance observed across the JJ, and $\Phi_0$ is the quantum of magnetic flux. The nonlinear relationship between phase and current in the JJ makes it an excellent candidate for nonlinear circuits and oscillators in different devices such as superconductor quantum interference devices (SQUIDs) and qubits \cite{ref25,ref26,ref27}.

\begin{figure}[ht]
\centering
\includegraphics[width=0.4\linewidth]{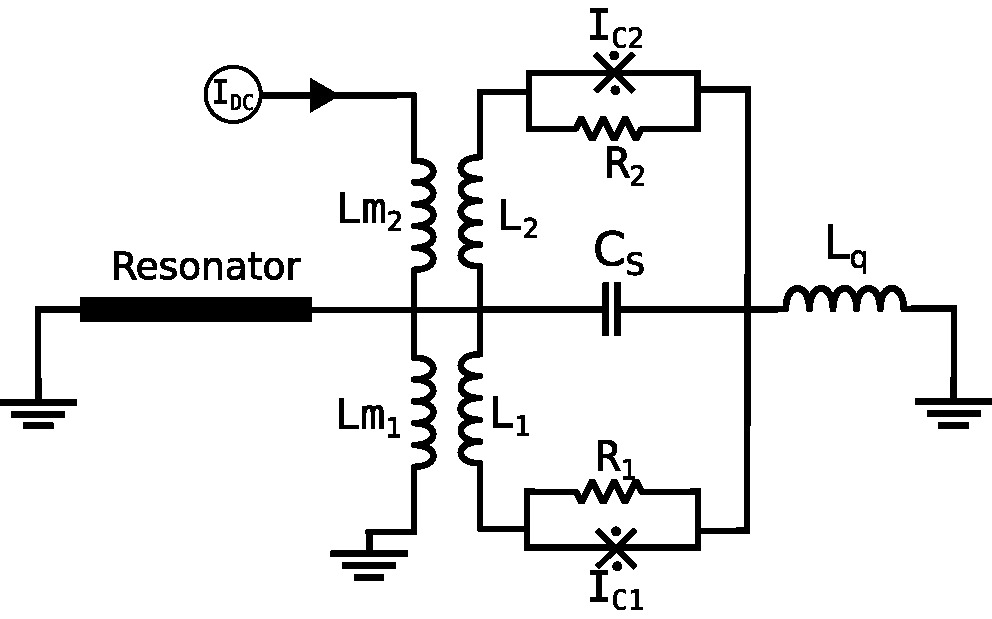}
\caption{Circuit schematic of the Josephson parametric oscillator used as a cell to emulate the spin resonance. For symmetry we assume that $L_1 = L_2 = 7.5pH$, $I_{C1} = I_{C2} = 80\mu A$ with 15 $\Omega$ shunt resistances, and the $C_S = 4.5pF$. The resonator frequency is at 7.5 GHz.}
\label{figure2.1}
\end{figure}

A SQUID is formed by connecting two JJs in parallel with a superconductor loop, as illustrated in Fig.~\ref{figure2.1}. The components of the SQUID include $L_1$, $L_2$, $I_{C1}$, and $I_{C2}$. The SQUID can be described as a single JJ whose critical current is $I_{C} = I_{C1} + I_{C2}$, and it is correlated with the applied flux to the SQUID loop. Therefore, the current of the SQUID can be modulated through magnetic coupling. In situations where the loop inductance value of the SQUID is small, the inductance value can be calculated as follows:
\begin{equation}
L_{SQUID} = \frac{\hbar}{2eI_{C}} \frac{1}{|\cos(\pi \Phi_{ext} /\Phi_0 )|},
\label{eq:4}
\end{equation}
$\Phi_{ext}$ represents the external magnetic flux in the SQUID loop, and $\varphi$ denotes the SQUID phase. In Josephson parametric oscillators (JPOs), we utilize the SQUID as a variable inductance, which can be controlled through external magnetic coupling. This capability enables the modulation of the oscillation frequency. Fig.~\ref{figure2.1} illustrates the resonator and SQUID loop combination, forming the JPO. The resonance frequency can be finely adjusted by altering the external flux on the SQUID loop.

\begin{figure}[ht]
\centering
\includegraphics[width=0.4\linewidth]{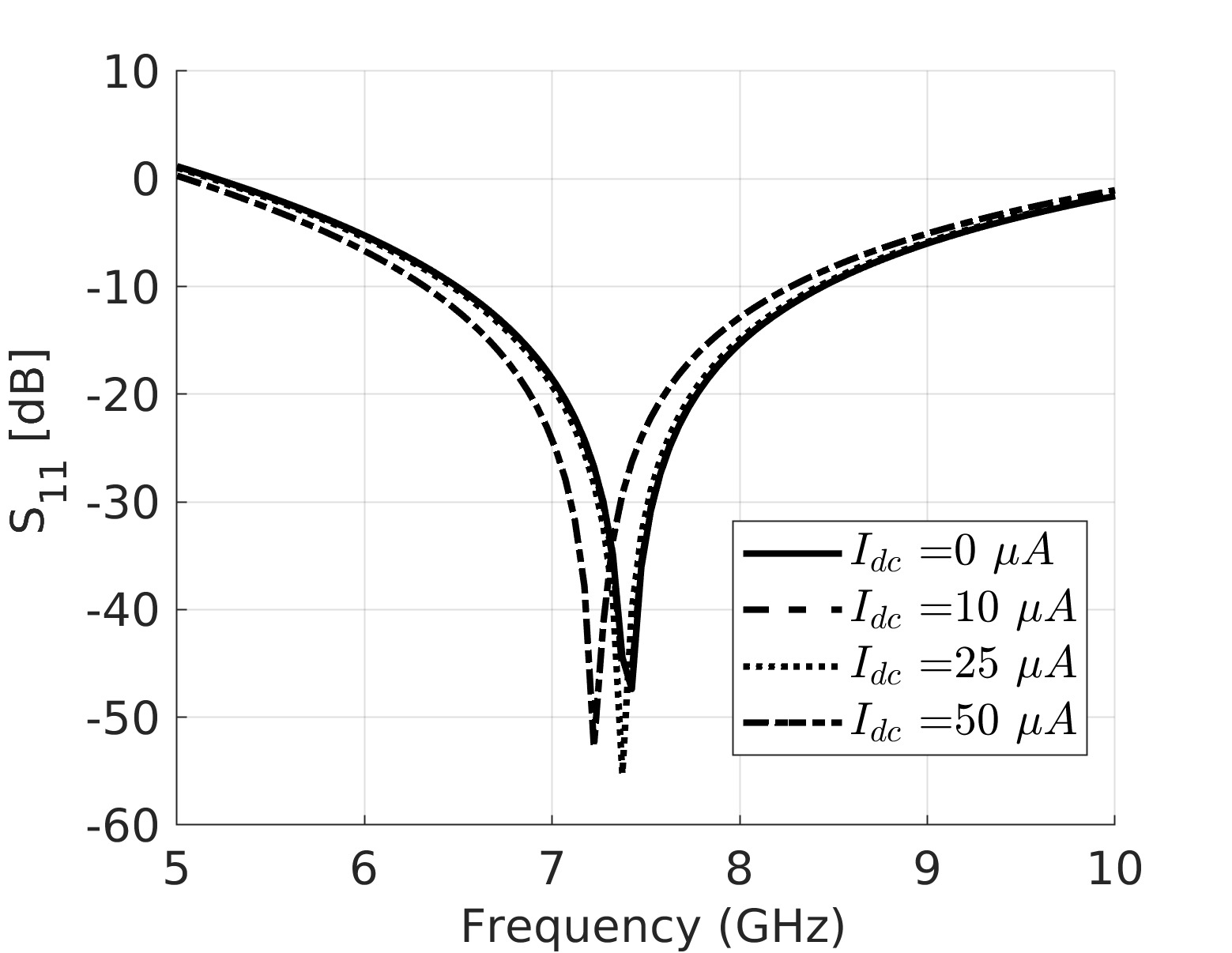}
\caption{Resonator frequency calculation result with different applied DC currents. The material is Nb, and the characteristics are chosen to match the MIT LL SFQee5 process.}
\label{figure2.2}
\end{figure}

Calculating the JPO's resonance frequency requires evaluating signal attenuation across various frequencies while considering the circuit as a simple quarter-wavelength resonator. Therefore, it becomes crucial to determine the impedance of the resonator. The surface impedance of a superconductor, also known as optical conductivity, can be estimated using Zimmerman's approximation of the Mattis-Bardeen equation \cite{ref28}. In these equations, the real part of impedance, $\sigma_1$, is due to quasi-particles, and the imaginary part, $\sigma_2$, corresponds to the super-current. The overall resonance frequency can be determined by combining the impedance of the resonator and the SQUID from Equation \ref{eq:4}. A DC flux would shift the inductance to match the frequency of oscillation. Therefore, the resonance frequency can be calculated as follows:
\begin{equation}
\omega_{0} = \omega_{r} \left[1+\frac{L_{SQUID}(\Phi_{ext})+L_{1}/2}{L_r}\right],
\label{eq:7}
\end{equation}
where $L_r$ represents the inductance of the resonator. In this work, we set $\omega_0$ to 7.5 GHz as the desired resonance frequency. Fig.~\ref{figure2.2} illustrates the variation of the resonance frequency when an external flux is applied to the SQUID loop through the $I_{dc}$ current source. In this figure, we calculated the oscillation frequency of the JPO, considering the parameters and limitations of the MIT LL SFQee5 process.
The parameter modulated in the JPO is the critical current of the SQUID. By coupling a small RF magnetic signal to the SQUID loop, the critical current, and consequently the inductance parameter, can be modulated, leading to a slight change in the resonance frequency. If the frequency of the pumped RF signal is twice the circuit's resonance frequency, in our case, 15GHz, it will deposit energy into the circuit until oscillation is sustained. Hence, the JPO is classified as a parametric oscillator. The JPO exhibits two stable points corresponding to the SQUID's 0 and $\pi$ phases. At these two phases, the flux in the SQUID is quantized to $\phi_0$, representing the state of minimum energy.

\subsection{Tile (Plaquette)}
\begin{figure}[ht]
\centering
\includegraphics[width=0.4\linewidth]{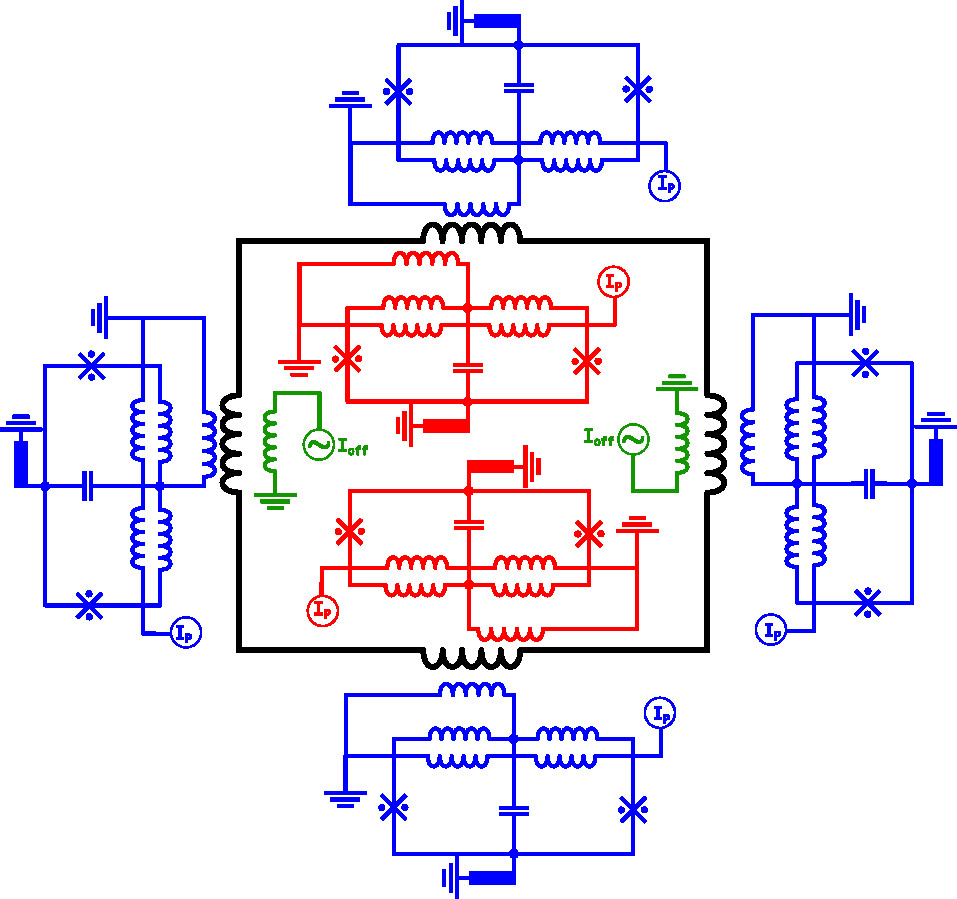}
\caption{A tile for the scalable architecture of IM. The structure has six JPOs, four as the logic and two as the ancilla. The coupler connects all JPO cells and provides interaction between them.}
\label{figure2.4}
\end{figure}

\noindent
The proposed tile employs coupled JPOs to realize spin interactions. The coupling between JPOs is achieved through magnetic coupling in an inductive loop, as depicted in Fig.~\ref{figure2.4}. The compact structure of the tile, compatible with commercial superconductor fabrication processes, features a coupler circuit highlighted in black. The coupler consists of a superconductor loop and a constant voltage source acting as an offset, which is pumped with the same frequency as the JPOs but at zero phase. Adjusting the pump signal allows the interaction between JPOs to be programmed in the IM.

An LHZ-compatible superconductive tile comprises six physical JPOs. Four are the main spins (JPOs), while the remaining two are ancillary bits. The physical coupling values of the main JPOs are identical, while the coupling values of the ancillary JPOs are twice as large. These coupling values can be modified by introducing a phase difference between the pump frequencies of the JPOs and the offset phase of the coupler loop. A phase difference of $0$ corresponds to fully coupled JPOs, while a phase difference of $\pi$ results in negative coupling. The tile structure solves combinatorial problems with up to four variables, while scalability is achieved by interacting multiple tiles based on the LHZ architecture. Considering each JPO as a spin, the total energy of this circuit can be expressed as,
\begin{equation}
E = \sum_{i=1}^{4} J_i\tilde{\sigma}_{i} - J_{a1}\tilde{\sigma}_{a1}\prod_{i=1}^{4} \tilde{\sigma}_{i} - J_{a2} \tilde\sigma_{a2}\prod_{i=1}^{4} \tilde\sigma_{i} - C_{cnst}\prod_{i=1}^{4} \tilde\sigma_{i}
\label{eq:plaquetteenergy}
\end{equation}
where $\sigma_i$ represents the logical JPOs, and $\sigma_{a1}$ and $\sigma_{a2}$ are the ancillary JPOs. $J_i$ represents the external field interaction, and $J_{a1}$ and $J_{a2}$ correspond to the ancillary interactions with the other JPOs. These values depend on the coupling values determined by physical coupling and the phase of the pump, as stated.
The constant value $C_{cnst}$ guarantees that the four-body interaction between the logical JPOs ($\prod_{i=1}^{4} \sigma_{i} $) is always positive.

The adjustment of parameter values in the superconductive tile can be achieved by modifying the phase of the pump's current source. It is crucial to satisfy this constraint in the context of the LHZ structure, where the number of spins with positive (negative) orientation must always be even, which is achieved by the ancillary JPOs, and the penalty term should always be negative, which is fulfilled by appropriately adjusting the parameter $C_{cnst}$ \cite{ref22}.

Furthermore, it should be noted that the ancillary JPOs are physically decoupled from each other and do not directly interact. When considering the external field effect on the logical JPOs ($J_b$) and the value of the ancillary JPOs ($J_a$), the expression for the plaquette energy (\ref{eq:plaquetteenergy}) can be rewritten as:
\begin{equation}
E = \sum_{i=1}^{4} J_b\tilde\sigma_{i} - (\tilde\sigma_{6} J_{a} + \tilde\sigma_{5} J_{a} + C_{cnst})\prod_{i=1}^{4} \tilde\sigma_{i}
\label{eq:plaquetteenergy2}
\end{equation}
The ancillary interaction described in this equation will result in ground energy values similar to equation (\ref{eq:LHZ_hamiltonian}), which characterizes the interactions in LHZ. Consequently, these ancillary interactions enable us to map the current pair-to-pair interaction circuit to the quadratic interaction of a tile required for the LHZ Hamiltonian implementation.
As mentioned earlier, altering the coupling values of the ancillary JPOs allows us to choose the expected ground states of the tile, enabling the mapping of different problems onto it. For instance, let's assume that all the logical JPOs have the same coupling value and the same interaction strength, denoted by $J_b$. In contrast, we assign random values to the ancillary interactions, $J_a$, that are physically twice as strong as the value of the logical interactions (i.e., $J_a = 2J_b$).
\begin{table}
\caption{\label{tab:table1}
The energy of different arrangements of spins for a sample coupling set tile. Here we consider the ancillary JPOs' physical coupling twice the logical JPOs.}
\begin{tabular}{ccc}
\hline
\shortstack{Physical \\JPO state}&\shortstack{Ancilla \\JPO state}& \shortstack{Total \\energy}\\
\hline
      $|1111\rangle$ & $|00\rangle$ & E = 4J$\textsubscript{b} - 2J\textsubscript{a} - J_{C} = -J_{C}$\\
      $|1110\rangle$ & $|00\rangle$ or $|11\rangle$ & $ E = 2J_{b} \pm 2J_{a} - J_{C}= 2J_{b} - J_{C}$\\
      $|1100\rangle$ & $|10\rangle$ & $E = -J_{C}$\\
      $|1000\rangle$ & $|11\rangle$ or $|00\rangle$ & $E = -2J_{b} \pm 2J_{a} - J_{C}= 2J_{b} - J_{C}$\\ 
      $|0000\rangle$ & $|11\rangle$ &  $E= -4J_{b} + 2J_{a} - J_{C} = -J_{C}$\\         
      \hline
\end{tabular}
\end{table}
Table \ref{tab:table1} illustrates the stable minimum energy states achieved in the circuit that depend on the values the ancillary JPOs settle into. When the ancillary settle into the $|00\rangle$ state, they force the physical JPOs to go to the $|1111\rangle$ state. When we set the ancillas to $|01\rangle$ or $|10\rangle$, only two logical JPOs can settle into the $|1\rangle$ state, and finally, in case the ancillas are settled at the $|11\rangle$ state, they force the physical JPOs to go to the $|0000\rangle$ state. This satisfies the ground state condition of the LHZ tile, where only even numbers of similar states should exist in a single tile.
\\

\section{\label{sec:Simulation}Evaluation and Discussion\protect\\}
\subsection{Analytical modeling of Ising Hamiltonian}
\noindent
To evaluate the functionality of the proposed tile circuit, we analytically modeled the Hamiltonian of a four-body interaction within the tile structure. For this, we have implemented equation \ref{eq:plaquetteenergy2} by considering noise using QuTip: Quantum toolbox in Python \cite{ref29}. In this case, the Hamiltonian formulation was   
\begin{equation}
 H = \sum_{i=1}^{4}J_i\sigma_{i}^{(z)} - C_{t} \prod_{i=1}^{4}\sigma_{i}^{(z)} + U_N
 \label{eq:HamiltonianCalc}
\end{equation}
where $J_i$ represent external field parameters, while $C_{t}$ represents the interaction arising from the coupler ($C_t = \sigma^x_5 \times J_{a} + \sigma^x_6 \times J_{a} + J_{,C}$), and the terms $\sigma^{(z)}$ correspond to Pauli's z matrix. Additionally, we incorporate $U_{N}$ as a noise component generated by multiplying a thermal coefficient by a pseudo-random energy distribution.

\begin{figure}[h]
\centering
\includegraphics[width=0.4\linewidth]{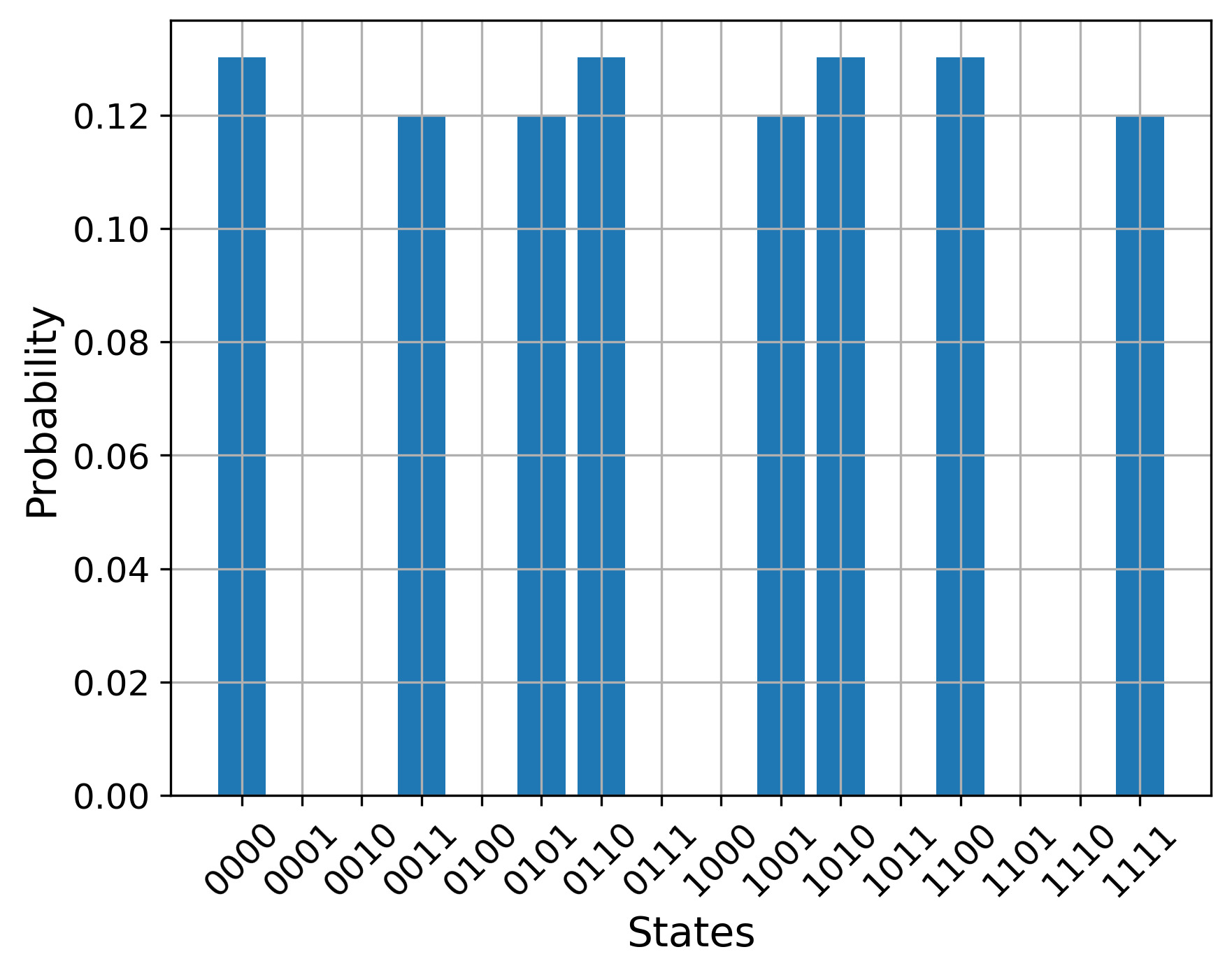}
\caption{Analytical modeling result of a tile that will be used for large-scale integration of IM. Here, the penalty term is positive, and the external field values are swiped so that the circuit can show all the possible states.}
\label{figure4.1}
\end{figure}

\begin{figure}[h]
\centering
\includegraphics[width=0.4\linewidth]{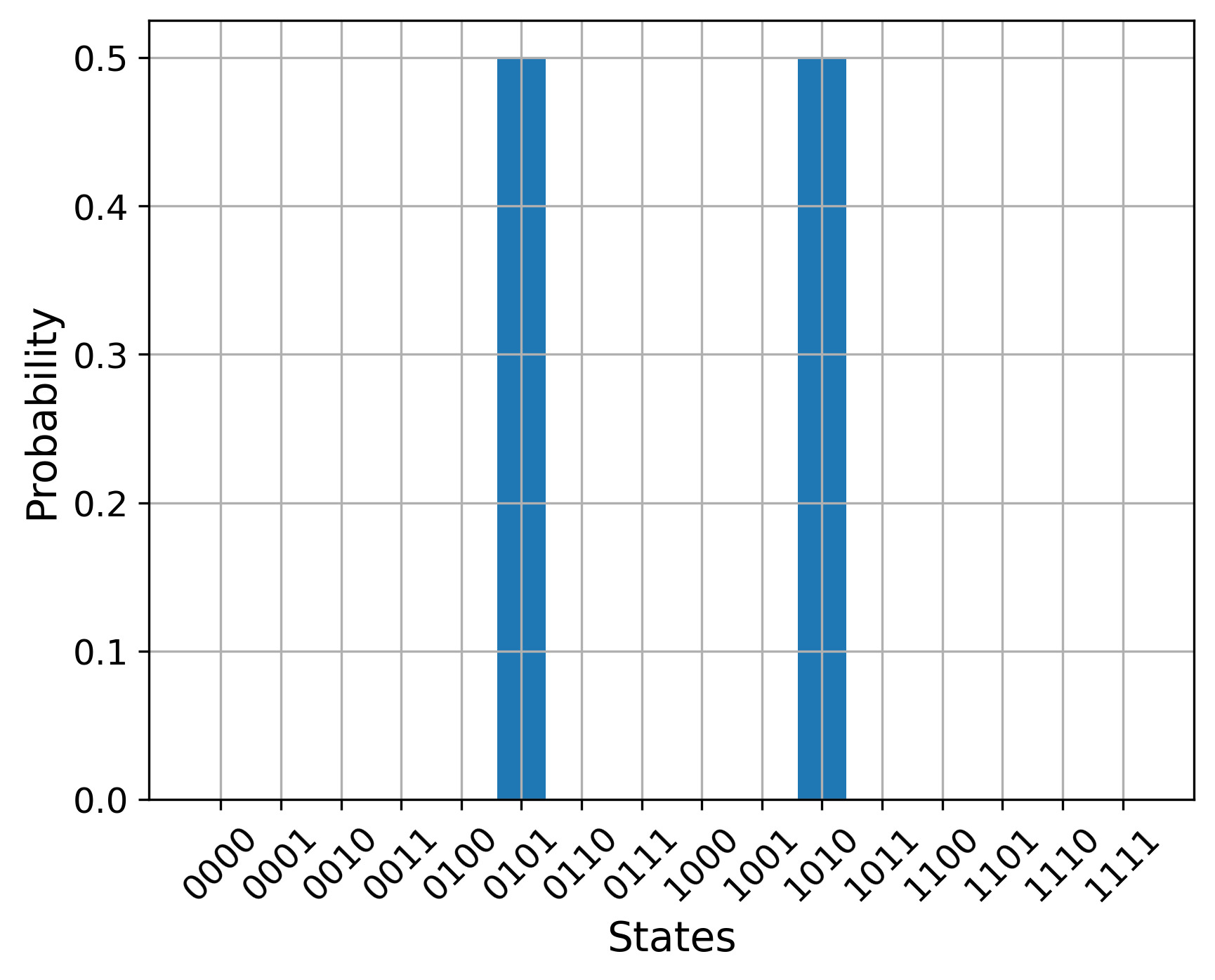}
\caption{Analytical modeling result of a tile that will be used for large-scale integration of IM. The external field values are fixed so that the circuit only settles on two of the minimum states.}
\label{figure3.5}
\end{figure}

The minimum energy states of the tile are investigated by introducing all the possible states and measuring the minimum energy value of the tile based on its Hamiltonian equation introduced in equation \ref{eq:HamiltonianCalc}. Then, based on the number of times the circuit is settled on a state, we calculate the distribution probability. The result for the probability of the states is shown in Fig.~\ref{figure4.1}. Here, the tile settles on the states that the number of 1's are even, as shown in Table \ref{tab:table1}.
When we set the external field values to a specific point, with half of the interactions being positive and half negative, the function settles only in two ground states as depicted in Fig.~\ref{figure3.5}. These two cases shown in Fig.~\ref{figure4.1} and Fig.~\ref{figure3.5} confirm that the solution of quantum Hamiltonian agrees with LHZ structure.  
\subsection{Analog simulation setup}
\noindent
We selected the MITLL SFQ5ee fabrication process for the proposed superconductor-based spin and tile. The MITLL process allows for a high current density of 100 $\mu$A/{$\mu$m}$^2$ for JJs. This process involves nine superconductor metal layers that facilitate the wiring and interconnect between JPOs and control circuits. Our design simulated a JPO with a resonance frequency of 7.5 GHz, which was pumped at 15 GHz. 

\begin{figure}[ht]
\centering
\includegraphics[width=0.4\linewidth]{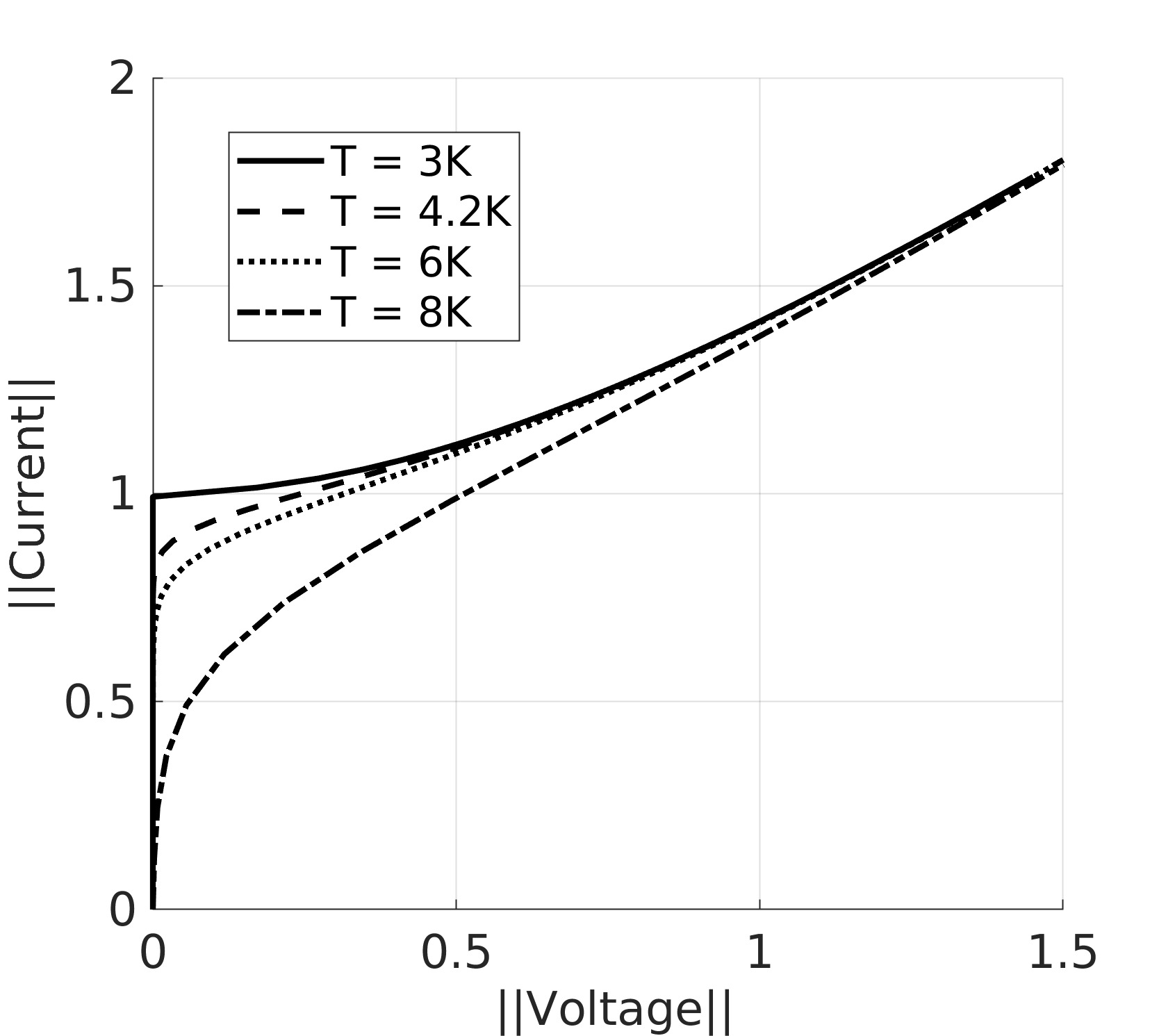}
\caption{Effect of Brownian noise on Josephson I-V characteristic.}
\label{figure2.3}
\end{figure}

We introduced thermal noise into the studied circuits to ensure more realistic simulation results. Thermal noise arises from the presence of normal electrons in normal metals and superconductors when the temperature exceeds absolute zero. This noise has two effects on superconductor circuits. The first effect, Johnson noise, can be modeled as white noise and applied to the circuit's resistances \cite{ref30}. The second effect of thermal noise in the circuit is due to interactions between Cooper pairs and normal electrons in the JJ. As the temperature increases, this noise causes slight variations in the JJ's current-voltage (I-V) characteristic and can be modeled as Brown noise \cite{ref31}. Fig.~\ref{figure2.3} illustrates the change in the I-V characteristic of our junction in the presence of Brown noise at different temperatures.
Semi-stochastic white noise sources were added paralleled to resistances in the circuit's netlist at the sources that apply bias and pump currents. Its value is given by $I_N = \sqrt{4RTK_B}$, where $T$ represents the noise temperature, $K_B$ denotes the Boltzmann constant, and $R$ signifies the resistance value.
To analyze the circuit's behavior and specifically observe its oscillation, we utilized the SPICE-based simulator JoSIM. We performed simulations and obtained time-domain results. Then fast Fourier transform (FFT) of the time-domain results was calculated to extract the frequency and phase characteristics. The state of the JPO is determined via its phase.

\subsection{Analog simulation results}
\begin{figure}[h]
\centering
\includegraphics[width=0.4\linewidth]{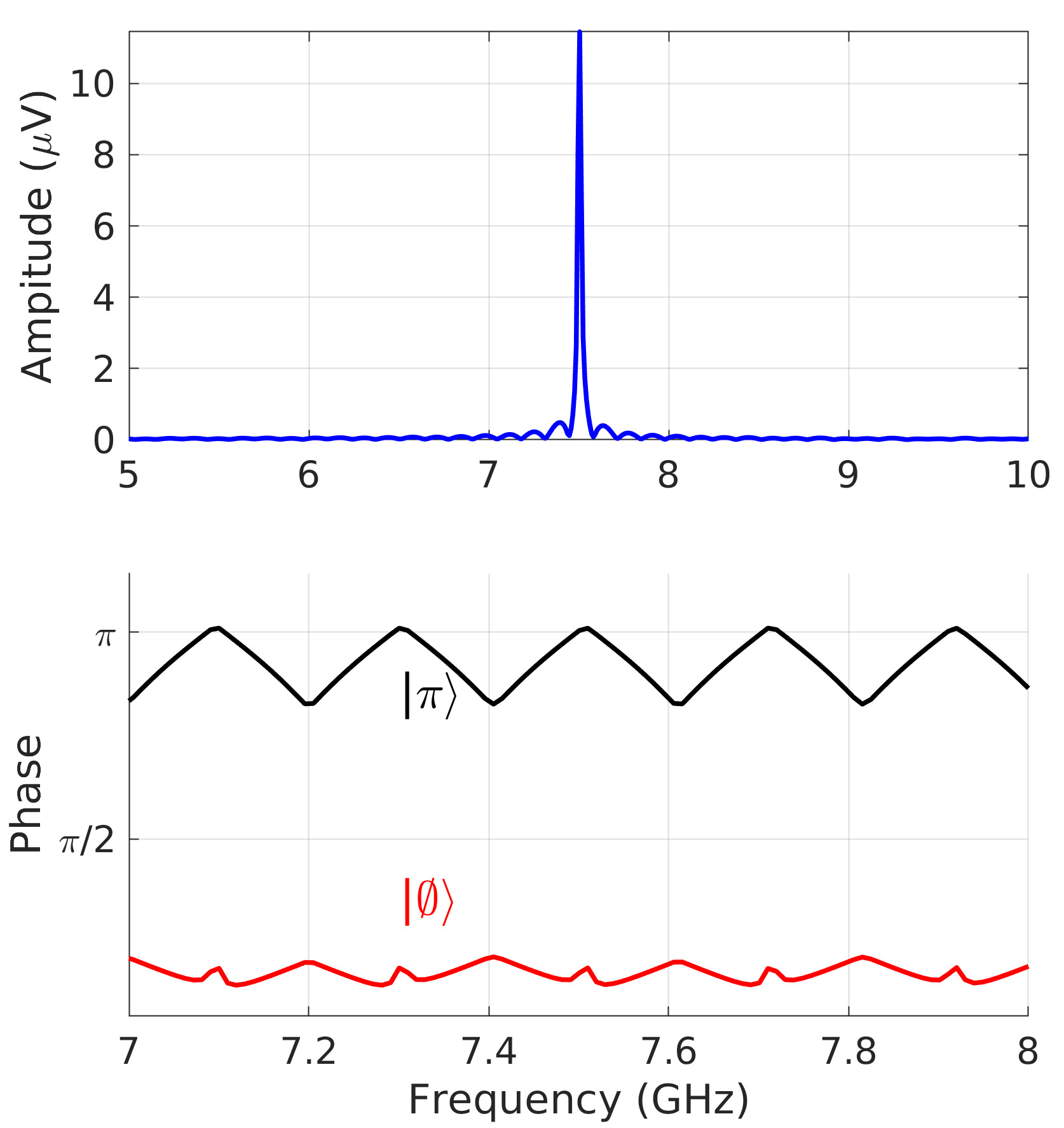}
\caption{Amplitude and Phase of the Simulated JPO Cell. Following simulation with JoSIM, Fast Fourier Transform (FFT) analysis reveals distinct phase states at the resonance frequency obtained from multiple simulations.}
\label{figure3.1}
\end{figure}

\noindent
Fig.~\ref{figure3.1} illustrates the FFT output of the SQUID loop's voltage obtained from simulating a single JPO cell. As shown, the resonance frequency of the JPO is precisely at 7.5GHz, half of the $I_{pump}$'s frequency. We ran multiple simulations with different random initial conditions. We observed that the measured phase settles in two distinct values, indicating the presence of two states with a phase difference of $\pi$.

It's important to acknowledge that numerical simulators, such as JoSIM, have certain limitations regarding time step, simulation duration, and the FFT algorithm. Consequently, the phase states observed may not be precisely $0$ and $\pi$, exhibiting slight variations. However, these variations are still distinguishable, identifying the two distinct phase states. This demonstrates the functionality of the simple JPO and establishes the $|\pi\rangle$ state as "1" and the $|0\rangle$ state as "0."
To change the interaction between two JPOs, we can manipulate the phase values in the pump signal. When two JPOs start with the same pump phase, constructive interference occurs, resulting in maximum coupling between them. Conversely, destructive interference occurs when there is a phase difference of $\pi$ between the JPOs, leading to minimal coupling. The same destructive and constructive patterns can be used in ancillary JPOs to change the penalty term of the tile.

\begin{figure}[ht]
\centering
\includegraphics[width=0.4\linewidth]{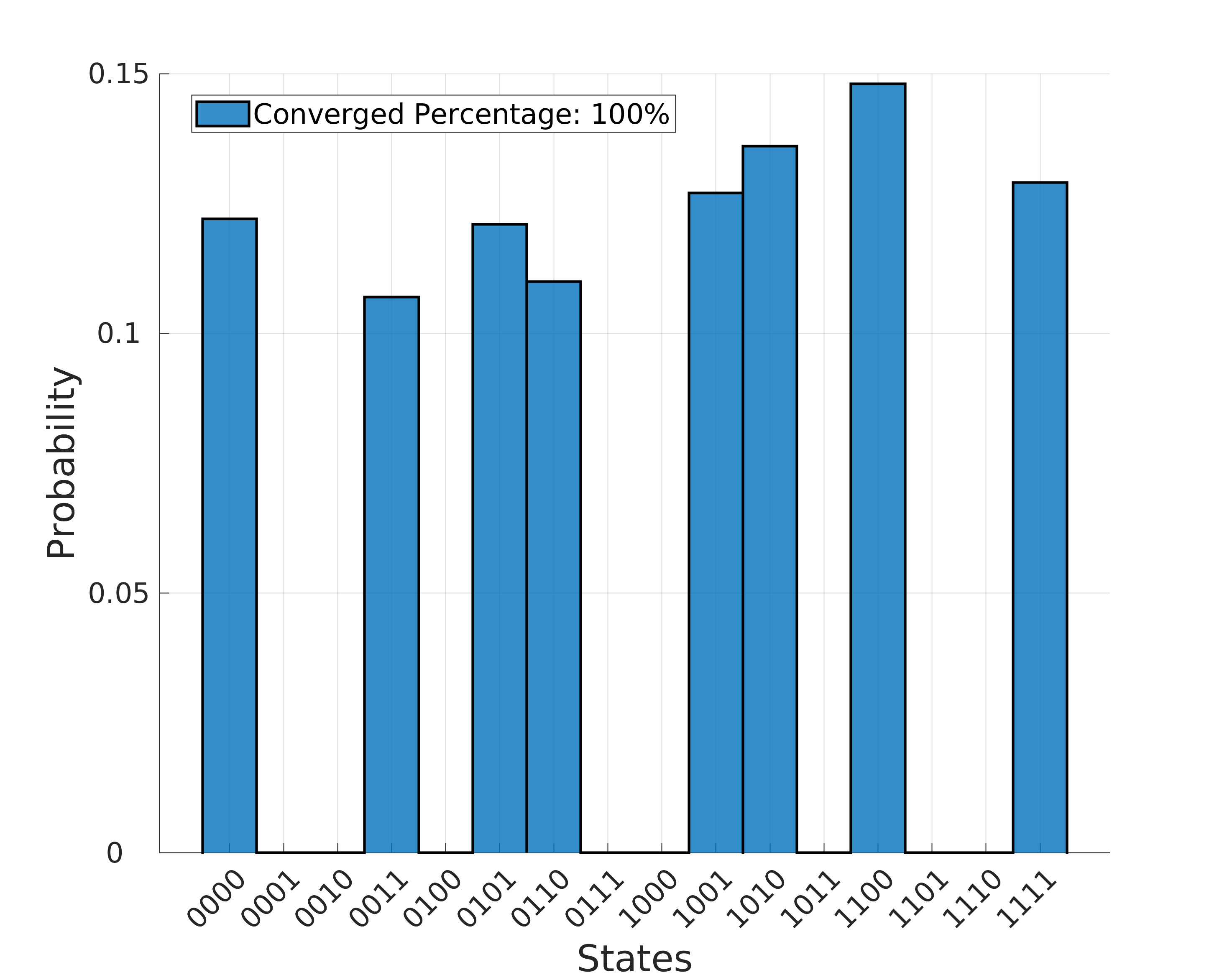}
\caption{Results for $1 k$ simulations on the 4-spin system with all the same interaction on a 4-node tile for $20 ns$ at random input points. The states we expect are at even parities. The results converge to the right answer in all of the cases in a noisy environment.}
\label{figure3.2}
\end{figure}

\begin{figure}[ht]
\centering
\includegraphics[width=0.4\linewidth]{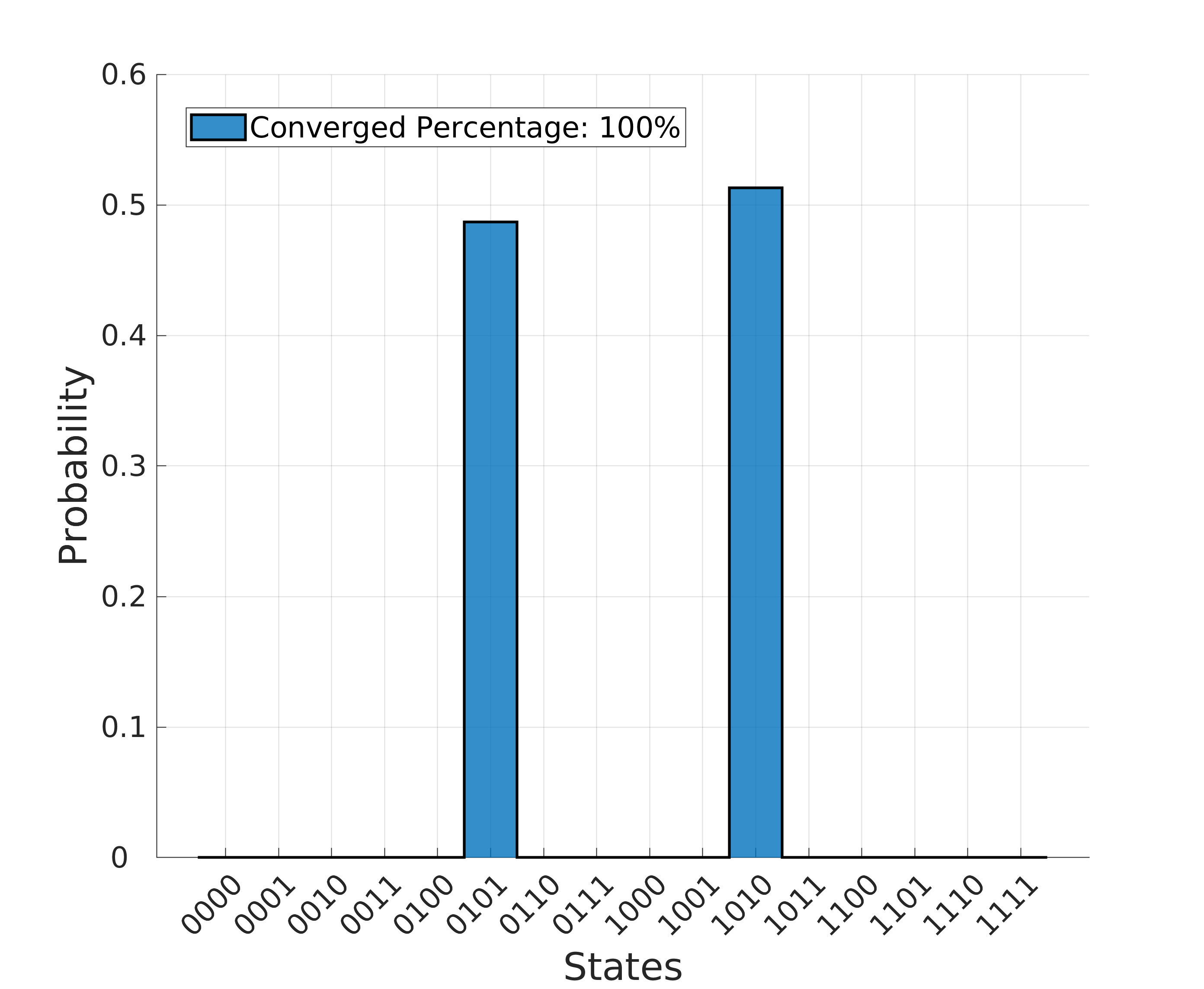}
\caption{Simulation result for $1 k$ run on the 4-parameter network with fixed phase values on JPOs on a 4-node complete graph ran for $20 ns$. The states we expect are $|0101\rangle$ and $|1010\rangle$.
}
\label{figure3.3}
\end{figure}

To validate the functionality of the tile, we implemented the tile design as depicted in Fig.~\ref{figure2.4} and compared the simulation results of the circuit with the Hamiltonian formulation. Initially, we simulated the tile by applying all possible states to the JPOs in the tile. In this scenario, the ancillary JPOs were set at $\frac{\pi}{2}$, and we initialized the JPOs with random phases selected from the set ${0, \pi}$. In the four-parameter problem, there exist 16 different states, and in this study, all the states with an even number of 1's are possible solutions (see Section \ref{sec:Theory} and the states shown in Table \ref{tab:table1}). We conducted 1k simulations; the outcomes are depicted in Fig.~\ref{figure3.2}. The expected solutions are the eight states with their respective probabilities shown. The slight differences in the probabilities of the states result from the added noise to the bias and pump sources, as well as the applied offset to the interaction. Considering the added noise, our current design can settle in all the ground states in all cases.
By setting the ancillary bits at half the phase and fixing the main JPOs phase values, we performed 1k simulations and observed the results illustrated in Fig.~\ref{figure3.3}. These simulation results align with the outcomes obtained when the ancillary JPOs were in the $|10\rangle$ state, as shown in Table \ref{tab:table1}. The results consistently converged to the correct states in all the runs. As depicted in Fig.~\ref{figure3.3}, the probabilities of obtaining each state as the answer were equal, and any slight differences in the probabilities can be attributed to the added noise. These findings agree with the emulation results illustrated in Fig.~\ref{figure3.5}.
%
\section{\label{sec:Results}Summary and Conclusion\protect\\}
\noindent
Superconductor electronics offer a potential solution for high-speed and low-energy computing compared to conventional devices. However, their unique properties have been largely overlooked in developing novel computing applications. In this study, we leveraged the nonlinear phase response of Josephson junctions to design and simulate a network of oscillators capable of solving Ising Hamiltonian problems.

The Josephson Parametric Oscillators (JPOs), operating at a frequency of 7.5 GHz, are interconnected through a coupler circuit. By utilizing four logical JPOs and two ancillary JPOs, we constructed a unit cell with quadratic interaction between JPOs known as a tile, which forms the building block of a large-scale Ising machine solver using the LHZ architecture.

To validate our design, we studied the Hamiltonian energies of the tile in two different scenarios and compared them with the numerical simulation results. We simulated it in a noisy environment, and the circuit stabilized in a few nanoseconds in all cases. The comparison between the calculation and numerical simulation showed a perfect match and satisfied the LHZ conditions. Therefore, the tile is suitable for implementation in the LHZ structure. Overall, our findings demonstrate the immense potential of superconductor electronics in computing, particularly for tackling computationally intensive problems through the utilization of Ising solvers based on Josephson junctions.
\section*{acknowledgments}
This work has been funded by the National Science Foundation (NSF) under the project Expedition: Discover (Design and Integration of Superconducting Computation for Ventures beyond Exascale Realization) grant number 2124453.\\
The authors thank Beyza Zeynep Ucpinar, Mustafa Altay Karamuftuoglu, and Yasemin Kopur for their help on this work.

\end{document}